\documentclass[aps,twocolumn, showpacs,superscriptaddress]{revtex4}
\usepackage{color}
\usepackage{graphics,graphicx}
\usepackage{amsmath}
\usepackage{tabularx}
\usepackage{footnote}
\begin{document}
\title{Proposal for Carrier-Envelope-Phase Stable Single-Cycle Pulse Generation in the Mid-Infrared--–Extreme-Ultraviolet Range}
\author{Z.\ Tibai}
\affiliation{Institute of Physics, University of P\'ecs, 7624 P\'ecs, Hungary}
\author{Gy.\ T\'oth}
\affiliation{Institute of Physics, University of P\'ecs, 7624 P\'ecs, Hungary}
\author{M.\ I.\ Mechler}
\affiliation{HAS-PTE High Field Terahertz Research Group, 7624 P\'ecs, Hungary}
\author{J.\ A.\ F\"ul\"op}
\affiliation{HAS-PTE High Field Terahertz Research Group, 7624 P\'ecs, Hungary}
\author{J.\ Hebling}
\affiliation{Institute of Physics, University of P\'ecs, 7624 P\'ecs, Hungary}
\affiliation{HAS-PTE High Field Terahertz Research Group, 7624 P\'ecs, Hungary}
\begin{abstract}
A robust method for producing half-cycle---few-cycle pulses in mid-infrared to extreme ultraviolet spectral ranges is proposed. It is based on coherent undulator radiation of relativistic ultrathin electron layers, which are produced by microbunching of ultrashort electron bunches by a TW power laser in a modulator undulator. According to our numerical calculations it is possible to generate as short as 10~nm long electron layers in a single-period modulator undulator having an undulator parameter of only $K=0.25$ and which is significantly shorter than the resonant period length. By using these electron layers the production of carrier-envelope-phase stable pulses with up to a few nJ energy and down to 30~nm wavelength and 70~as duration is predicted.
\end{abstract}
\maketitle

Waveform-controlled few-cycle laser pulses enabled the generation of isolated attosecond pulses and their application to the study of electron dynamics in atoms, molecules, and solids \cite{Krausz}. Intense \textit{waveform-controlled} extreme ultraviolet (EUV)/X-ray attosecond pulses could enable precision time-resolved studies of core-electron processes by using e.g. pump-probe techniques \cite{Tzallas}. Examples are time-resolved imaging of isomerisation dynamics, nonlinear inner-shell interactions, or multi-photon processes of core electrons. EUV pump--–EUV probe experiments can be carried out at free-electron lasers (FELs) \cite{Jiang,Magrakvelidze}; however, the temporal resolution is limited to the fs~regime and the stochastic pulse shape is disadvantageous.

The shortest electromagnetic pulses reported to date, down to a duration of only 67~as, were generated by high-order harmonic generation (HHG) in gas targets \cite{Goulielmakis,Zhao}. Isolated single-cycle 130-as pulses were generated by HHG using driving pulses with a modulated polarization state \cite{Sansone}. One drawback of gas HHG is the relatively low EUV pulse energy due to the ionization depletion of the medium. The use of long focal length for the IR driving field, or using strong THz fields for HHG enhancement \cite{Kovacs} were proposed to increase the EUV pulse energy. The generation of half-cycle 50-as EUV pulses with up to 0.1~mJ energy is predicted by coherent Thomson backscattering from a laser-driven relativistic ultrathin electron layer by irradiating a double-foil target with intense few-cycle laser pulses at oblique incidence \cite{WuPRL10,WuNatPh12}. Various schemes, such as the longitudinal space charge amplifier \citep{Dohlus,Marinelli}, or two-color enhanced self-amplified spontaneous emission (SASE) \cite{Zholents2,Ding2} were proposed for attosecond pulse generation at FELs. However, the realization of these technically challenging schemes has yet to be demonstrated and precise waveform control is difficult.

In this Letter we propose a robust method for producing waveform-controlled pulses down to half-cycle durations in the mid-infrared (MIR) to the EUV spectral ranges. The method is based on coherent undulator radiation emitted by relativistic ultrathin electron layers, which are produced by microbunching of a picosecond electron bunch obtained from microwave electron injectors/accelerators in a modulator undulator driven by a few-TW visible laser.

Evidently, efficient generation of radiation by coherent scattering is possible only if the (micro)bunch length is shorter than the radiation's half period. Although production and compression of microbunches by using a modulator undulator were reported, the shortest microbunch length was about 800~nm \cite{KimuraPRL01}. In that experiment electron bunches with a relativistic factor of $\gamma = 89$ were microbunched by using the 10.6-$\mu$m radiation of a high-power CO$_2$ laser. If the Coulomb interaction between the electrons of the bunch can be neglected an analytical formula can be derived for the distribution of the electron density of the microbunched structure \cite{ZholentsPRST05}. According to this formula the $\Delta z_0$ minimal width of a microbunch is given by
\begin{equation}
\Delta z_0=\frac{\lambda_l}{2B},\label{eq:Dz0}
\end{equation}
where $\lambda_l$ is the wavelength of the laser, and $B=\Delta\gamma_w/\sigma_{\gamma_0}$ is the ratio of the induced change of the electron relativistic factor by the microbunching process and the scattering of the relativistic factor in the original electron bunch. According to Eq.~(\ref{eq:Dz0}), shorter microbunches can be achieved by using shorter laser wavelength, more monoenergetic electron beam and stronger energy modulation.

We used the GPT numerical code \cite{GPT} for the simulation of microbunching by the modulator undulator. This code takes into account space-charge effects by using the model described in \cite{Poplau}. As it is usual, in order to reduce the calculation time, macroparticles were considered rather than individual electrons. In the first simulations most electron bunch parameters before the microbunching were chosen according to those of an existing injector \cite{YangJJAP05}. These parameters and the most important parameters of the laser and the modulator undulator are listed in Table~\ref{tab:1}.
\begin{table}
\caption{\label{tab:1}{Parameters and their values used in our first series of calculations.}}
\begin{tabularx}{\columnwidth}{lXc}
\hline
\hline
\multicolumn{1}{c}{Parameter} && Value\\
\hline
E-beam energy && 31.7 MeV\\
E-beam intrinsic energy spread (1 $\sigma$) && 0.04\% \\
E-beam charge (total pulse) && $\approx$ 1.2~nC\footnote{This value is by 20\% higher than the value of Ref. \cite{YangJJAP05}.}  \\
E-beam pulse length (1 $\sigma$) && $\approx$ 1.8 ps\\
E-beam normalized emittance && 3.2 mm mrad\\
E-beam radius && 140 $\mu$m\\
Undulator period length && 9.7~mm\\
Laser wavelength && 1.3~$\mu$m\\
Laser power && 3.89 TW\\
Laser beam size inside the modulator u. && 0.72 mm\\
\hline
\hline
\end{tabularx}
\end{table}
The $\lambda_u$ undulator period was chosen so that it satisfied the well known
\begin{equation}
\lambda_u=\frac{2\gamma^2\lambda_l}{1+K^2/2}\label{eq:rescon}
\end{equation}
resonance condition \cite{Schmuser}.

\begin{figure}
\includegraphics[trim =30 30 30 20,width=\columnwidth]{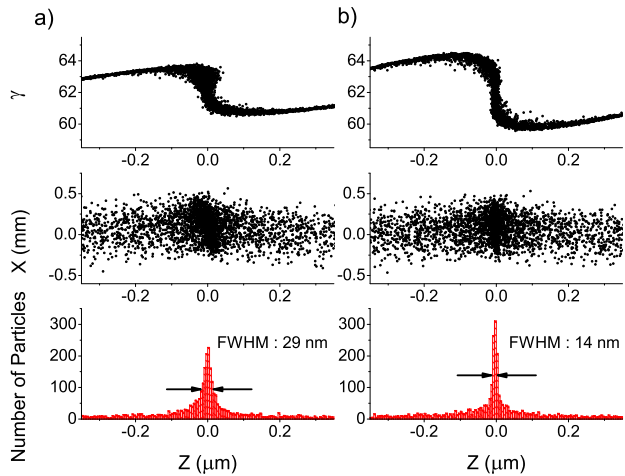}
\caption{\label{fig:1} Simulated distribution snapshots of quasi-particles after compression by a single-period modulator undulator with a "resonant" undulator period length $\lambda_u$ as given by Eq.~(\ref{eq:rescon}) (a) and with an optimal undulator period length (b). The three panel rows show the distribution of the relativistic factor $\gamma$ along the $z$~axis, the spatial distribution in the $x$-$z$~plane, and the spatial distribution along the $z$~axis, respectively.}
\end{figure}

Inside the modulator undulator the interaction between the electrons, the magnetic field of the undulator and the electromagnetic field of the laser introduces a periodic energy modulation of the electrons along the bunch with a period of the laser wavelength \cite{KimuraPRL01}. This energy modulation leads to the formation of microbunches in the drift space behind the undulator. For the parameters used the shortest microbunch duration was reached at $41$~mm behind the centre of the undulator. Fig.~\ref{fig:1}
displays the energy and spatial distributions of the electrons around this point. The microbunch length is as short as 29~nm, which is 28~times shorter than predicted in Ref.~\cite{KimuraPRL01}. From this difference the assumed shorter laser wavelength is responsible for a factor of 8 according to Eq.~(\ref{eq:Dz0}). The assumed high laser power (Table \ref{tab:1}) in our simulation results in a larger $B$ value causing a stronger microbunch shortening according to Eq.~(\ref{eq:Dz0}). Furthermore, it is important to notice that we obtained the shortest microbunches for $K\approx 0.25$. This value is one order of magnitude smaller than the $K=3$ value used in \cite{KimuraPRL01}. At $K=3$ our calculations predict 16~times longer microbunches than at $K=0.25$.

Another interesting and somewhat surprising result of our simulations is that a shorter microbunch length can be achieved by using an undulator with a period shorter than the $\lambda_u$ resonant period obtained from Eq.~(\ref{eq:rescon}). As an example, Fig.~\ref{fig:1}b displays the calculated energy- and spatial distributions of the electrons when all of the parameters in the simulations were the same as used for Fig.~\ref{fig:1}a, except that now 5.3~mm long undulator period was assumed instead of 9.7~mm. The shorter undulator period resulted in 14~nm long microbunches, which is two times shorter than in case of the longer period. Comparison of the upper panels of Figs.~\ref{fig:1}a and \ref{fig:1}b indicates that the main reason of the shorter microbunch for shorter undulator is the stronger modulation of the electron energy. A two-term expression is obtained for the electron energy modulation introduced inside an undulator \cite{Schmuser}. One term describes a fast oscillation with the interaction length. In the usual case of multi-period undulators this term averages out and the resonant condition (Eq.~(\ref{eq:rescon})) is obtained from the remaining other term. The reason for obtaining the largest electron modulation at an undulator period shorter than predicted by Eq.~(\ref{eq:rescon}) is obviously the fact that in our case of a single-period undulator none of the two terms is negligible.

The 14~nm microbunch length is as much as 93~times shorter than the wavelength of the laser used for the microbunching. Still, this value is about 1.5~times larger than predicted by Eq.~(\ref{eq:Dz0}) and the $\Delta\gamma_w=1.8$ value obtained from the upper panel of Fig.~\ref{fig:1}b. This difference is caused by the Coulomb interaction, which was neglected during the derivation of Eq.~(\ref{eq:Dz0}).

Coherent synchrotron radiation in the far-infrared region was considered 30~years ago \cite{MichelPRL82}. Recently both broadband \cite{CarrNat02} and narrowband \cite{ShenPRL11} efficient THz sources based on this effect were demonstrated. In these publications the experimental results were explained by using an expression giving the energy radiated per unit solid angle per unit frequency interval \cite{CarrNat02,ShibataPRA92}.
Since we are interested in the temporal shape of the ultrashort pulses emitted by the extremely short electron microbunches in the radiator undulator we calculated it in a plane 8~m behind the middle of the radiator undulator according to \cite{Jackson}:
\begin{equation}
\vec{E}(\vec{r},t)=\left[ \frac{q \mu_0}{4 \pi} \frac{\vec{R} \times \left((\vec{R}-R\vec{\beta}) \times {\dot{\vec{v}}}\right)}{(R-\vec{R} \cdot \vec{\beta})^3} \right]_{\text{ret}},\label{eq:vecE}
\end{equation}
where $q$ is the electric charge of a macroparticle, $\mu_0$ is the vacuum permeability, $\vec{R}$ is the vector pointing from the position of the charge at the retarded moment to the observation point, $\vec{v}$ is the velocity of the macroparticle, and $\vec{\beta}=\vec{v}/c$, where $c$ is the speed of light, and $N$ is the number of the macroparticles. During the radiation process the acceleration, velocity, and position of the macroparticles were traced numerically by taking into account the Lorentz force equation:
\begin{equation}
\frac{\mathrm{d}(\gamma m \vec{v})}{\mathrm{d}t}=q\vec{v} \times \vec{B}\left( z(t)\right),\label{eq:dgamma}
\end{equation}
where $\vec{B}$ is the undulator magnetic field at the position of the macroparticle. The Coulomb interaction between the macroparticles was neglected during the radiation process. This simplification is justified by the short interaction time. 

In a series of calculations electron bunches with $\gamma = 900$, transversal size of 80~$\mu$m, and undulators with a few different magnetic field distributions were assumed. The results of these calculations indicate the versatility of the proposed method for the generation of ultrashort pulses with pre-defined waveforms. Pulses with a few oscillation cycles and relatively narrow spectra can be generated by an undulator consisting of a few magnetic dipole pairs. The number of oscillations decreases and the spectrum becomes broader when the number of magnet pairs is decreased. The carrier-envelope phase \cite{carrier} can also be controlled by the undulator's magnetic field distribution for our attosecond pulses.

\begin{figure}
\includegraphics[trim =30 70 30 70,width=\columnwidth]{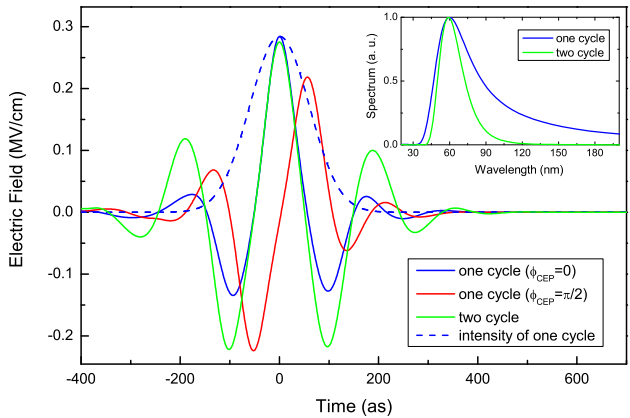}
\caption{\label{fig:2} Examples of calculated pulse shapes and spectra (inset) for a few different magnetic field distributions and $\gamma = 900$. }
\end{figure}

\begin{figure}
\includegraphics[trim =40 50 50 50,width=\columnwidth]{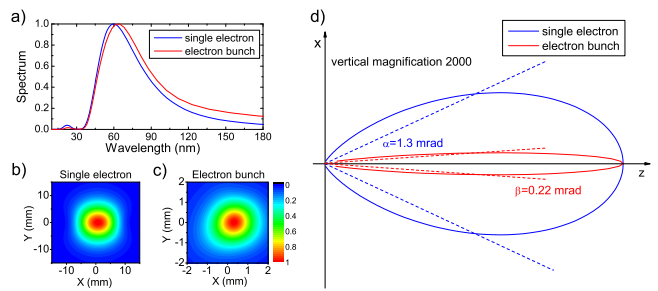}
\caption{\label{fig:3}Comparison of single-electron and bunch results for single-cycle pulses and $\gamma=700$. The radiation wavelength is much longer than the bunch length. Details are given in the text.}
\end{figure}

Besides the much shorter microbunches used in our simulations as compared to previous works using undulator radiation for THz generation \cite{CarrNat02,ShenPRL11}, there is another important difference between the two cases. For the microbunches used in Refs.~\cite{CarrNat02} and \cite{ShenPRL11} the transversal size was much smaller than their longitudinal size. Because of this, the properties of the generated radiation were solely determined by the longitudinal distribution of the electrons inside the bunch, the effect of the transversal distribution was negligible. Contrary to this, in our case the transversal size, being on the order of $100~\mu$m, was more than three orders of magnitude larger than the longitudinal size of the microbunch. This means that although the longitudinal size of the microbunch is significantly shorter than half of the $\lambda_r$ radiation wavelength calculated from the
\begin{equation}
\lambda_r=\frac{1+{K^2}/2}{2 \gamma^2 \lambda_u}\label{eq:lambdar}
\end{equation}
formula, the transversal size is a few orders of magnitude larger than $\lambda_r$. One consequence of this is shown in Fig~\ref{fig:3}a. The spectral peak of the generated radiation is shifted to a longer wavelength as compared both to the peak predicted by Eq.~(\ref{eq:lambdar}) as well as to that obtained by considering only a single particle in the calculation. The explanation of this effect is very simple. For large transversal size, even for an observation point on the axis determined by the center and propagation direction of the electron bunch, the difference of the distances from the observation point to a particle being close and to a particle being far from the axis can be comparable to the radiation wavelength. This causes destructive interference suppressing high-frequency components in the generated radiation, thereby shifting the spectrum to longer wavelengths.

The large microbunch transverse size also reduces the emission solid angle as compared to the radiation of a single electron. This reduction is illustrated by Fig.~\ref{fig:3}b-d. For a relativistic factor of $\gamma = 700$ the divergence angle of the radiation measured at the intensity half maximum is almost 6~times smaller for a bunch with $80~\mu$m radius than for a single electron. The reason for this is that for off-axis observation points the distance to electrons at opposite transversal edges of the microbunch is different, and the difference increases with off-axis angle. This results in destructive interference limiting the radiation emission angle of bunches significantly below that of a single particle.

Other consequence of the destructive interference is that the energy of the radiation does not scale with the square of the number of macroparticles, as in case of negligible transversal and longitudinal extensions of the microbunch. For example, for a single macroparticle (with the charge of 4500~electrons but without any spatial extension) the calculated energy of the radiation (distributed according to Fig.~\ref{fig:3}b) is 80~fJ. For a microbunch with 330 such macroparticles within its 9-nm long FWHM region the calculated energy of the radiation (distributed according to Fig.~\ref{fig:3}c) is 470~pJ. The latter value is only 5875~times larger than the former, rather than $330^2~ \approx ~110000$~times which would follow from the square-law. Since the longitudinal extension of the bunch considered is more than six times shorter than the mean radiation wavelength, the $110000/5875 \approx 19$~times difference between the expected and observed energy increase is mainly caused by the relatively large transversal size of the microbunch. Indeed, decreasing the transversal size of the microbunch resulted in significant increase of the energy of the generated as pulse.

\begin{figure}
\includegraphics[width=\columnwidth,trim =10 50 10 10]{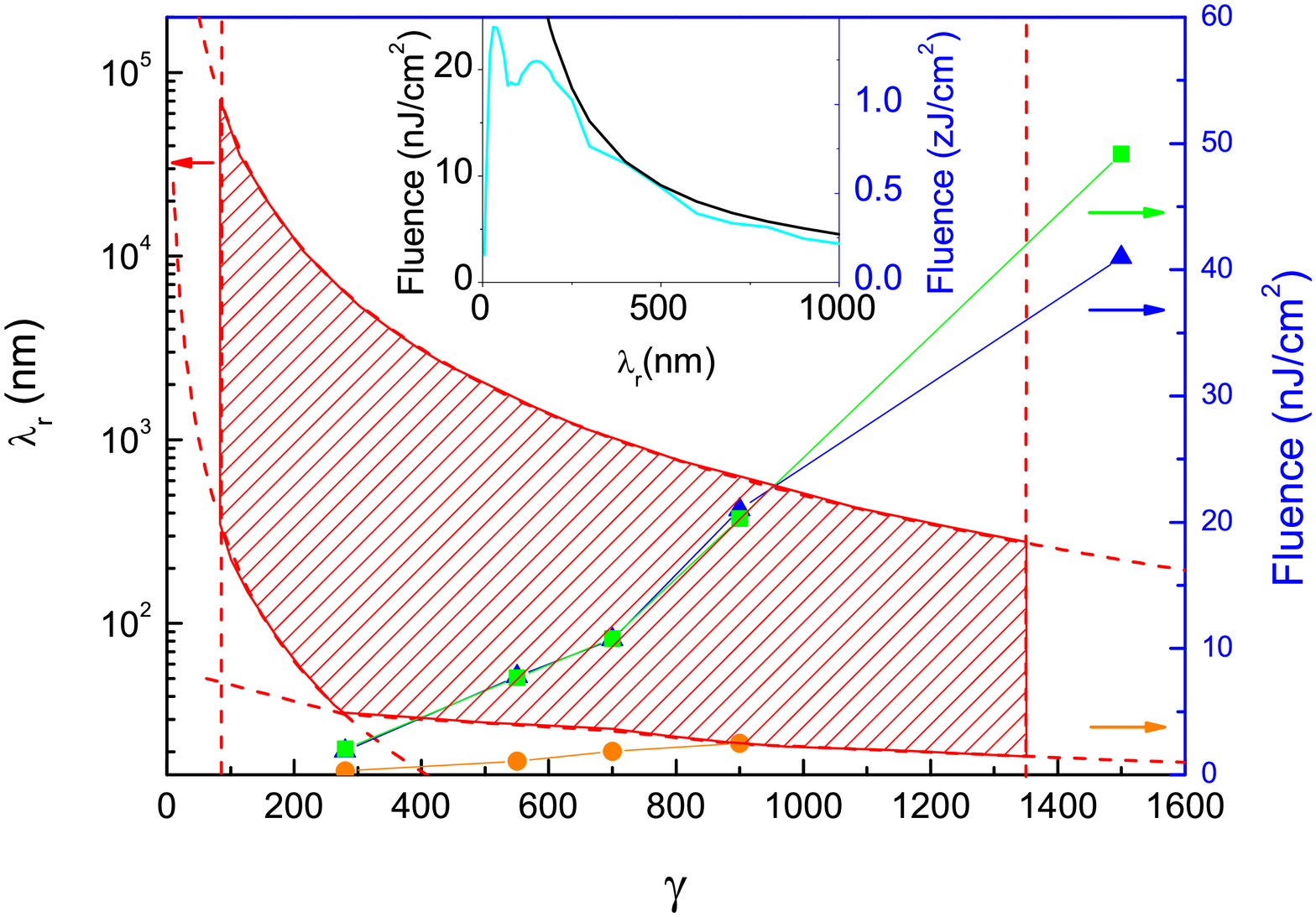}
\caption{\label{fig:4} Dependence of the pulse fluence on $\gamma$ for 30~nm (blue triangles) 60~nm (green squares) and 1~$\mu$m (orange dots). The shaded area indicates the available radiation wavelength and possible $\gamma$ ranges determined by restricting the period length of both undulators to the experimentally feasible 5~mm to 1000~mm range and by restricting the radiation wavelength to longer than two bunch lengths. The inset shows the calculated pulse fluence versus the radiation wavelength for a microbunch and for a single electron, both for $\gamma=900$.}
\end{figure}

There are two possibilities for decreasing the difference between the solid angles of the radiation emission of a single electron and the microbunch, and increasing in this way the energy of the pulse emitted by the microbunch. One possibility is reducing the transversal size of the microbunch. However, this requires spatial focusing. Furthermore, decreasing the transversal size at constant bunch charge results in larger Coulomb repulsion and, as a consequence, in longer microbunches prohibiting the generation of short wavelengths. Because of these complications, in this short Letter we do not investigate this possibility. The other possibility is increasing the relativistic  $\gamma$ factor of the electrons. It is well known that the number of photons emitted per magnet period is independent of $\gamma$ and it is given as \cite{Jackson}:
\begin{equation}
N=\frac{2\pi}{3}\alpha K^2,\label{eq:N}
\end{equation}
where $\alpha$ is the fine structure constant. It is also known that the solid angle of the radiation emitted by an electron moving with relativistic speed is proportional to $\gamma^{-2}$. Because of these facts it is expected that the fluence of the pulse emitted by a microbunch is proportional to $\gamma^2$ if the wavelength of the emitted radiation is kept constant.

Fig.~\ref{fig:4} displays results of calculations in which the undulator parameter was kept constant at $K=0.5$, the relativistic factor was varied in the range of $\gamma=280 \div 1500$, and for each $\gamma$ value the $\lambda_u$ undulator period was chosen such that the radiation wavelength given by Eq.~(\ref{eq:lambdar}) becomes $\lambda_r = 30$~nm, 60~nm and 1~$\mu$m, respectively. At small and medium values of $\gamma$ the fluence curves indeed indicate square dependence. At large $\gamma$ the fluence is significantly smaller than expected from the square dependence. The reason for this is that for large $\gamma$ bunch development needs as long as a few meters of drift space whereby the transverse size of the bunch increases significantly (up to about 0.5~mm for $\gamma=1500$) favouring destructive interference.

Although the fluence of the generated pulse at 1~$\mu$m is significantly smaller than for EUV wavelengths, the pulse energy is larger for the longer wavelength, because of the larger radiation solid angle. By using electron bunches with $\gamma=900$ single-cycle pulses with 0.3~nJ, 0.7~nJ, and 1.8~nJ energy can by generated at 34~nm, 60~nm, and 1~$\mu$m, respectively. This energy is high enough to use these pulses as pump in pump-probe measurements. Pulses with the predicted exceptional parameters can enable time- and CEP- resolved measurements with sub-100-as resolution. However, applications of such a set-up can be limited if the high-energy electron bunches are produced in common microwave accelerators, because of its high expenses. Laser-wakefield accelerators are a promising alternative source for high-energy electron bunches \cite{Leemans}. At present, the relatively large -- of a few percentage -- energy spread of the generated bunches prohibits their direct application. However, improvement in this respect is expected; furthermore, the slice emittance is the important property. Maybe it is possible to decrease the slice emittance significantly after a proper drift distance, or by using appropriate chicane. A detailed investigation of this possibility is needed.

Since in our proposed setup the electron bunch consists of many microbunches separated by the modulation laser wavelength, a pulse sequence is generated. The ratio of separation time to pulse duration is smaller than 20. One possibility we have to consider for increasing this ratio significantly is using two modulation lasers with substantially different wavelength in the modulator undulator \cite{Ding2}. This might also increase the charge of the microbunch and the energy of the generated pulses. However, investigation of this possibility and the combination of our scheme with other ones, such as the longitudinal space charge amplifier \cite{Marinelli,Dohlus}, or SASE FEL is out of the scope of this short article.

\end{document}